\documentclass[12pt]{article}
\usepackage{graphicx}
\usepackage{graphics}
\title{Addison-type series representation for the Stieltjes constants} 
\author{Mark W. Coffey\\
Department of Physics\\
Colorado School of Mines\\
Golden, CO  80401\\
(Received $\mbox{~~~~~~~~~~~~~~~~~~~~~~~~~~~~~~~2009}$)}
\date{December 6, 2009}  
\begin{document}
\maketitle
\baselineskip=25 pt
\begin{abstract}

The Stieltjes constants $\gamma_k(a)$ appear in the coefficients in the regular
part of the Laurent expansion of the Hurwitz zeta function $\zeta(s,a)$ about its 
only pole at $s=1$.  We generalize a technique of Addison for the Euler constant
$\gamma=\gamma_0(1)$ to show its application to finding series representations for these constants.  Other generalizations of representations of $\gamma$ are given.

\end{abstract}
 
\medskip
\baselineskip=15pt
\centerline{\bf Key words and phrases}
\medskip 

\noindent

Stieltjes constants, Riemann zeta function, Laurent expansion, series representation,
Hurwitz zeta function 

\vfill
\centerline{\bf 2010 AMS codes} 
11M06, 11Y60, 11M35

\baselineskip=25pt
\pagebreak
\medskip
\centerline{\bf Introduction and statement of results}
\medskip

The Riemann zeta function has but one simple pole, at $s=1$ in the complex
plane \cite{edwards,riemann,titch}.  In the Laurent series about that point,
$$\zeta(s)={1 \over {s-1}}+\sum_{k=0}^\infty {{(-1)^k \gamma_k} \over
k!} (s-1)^k, \eqno(1.1)$$
$\gamma_k$ are called the Stieltjes constants 
\cite{briggs,coffeyjmaa,hardy,kluyver,matsuoka,matsuoka2,mitrovic,stieltjes,wilton2},
where $\gamma_0=\gamma$, the Euler constant.  These constants have many 
uses in analytic number theory, asymptotic analysis, and elsewhere.  Among other applications, estimates for $\gamma_n$ may be used to determine a zero-free region of
the zeta function near the real axis in the critical strip $0 < \mbox{Re}
~s < 1$.  
The Stieltjes constants can be related to sums over the complex zeta function zeros,
as well as to the Li/Keiper constants (e.g., \cite{coffeyprsa}).

The Hurwitz zeta function $\zeta(s,a)$ may also be analytically continued to the 
whole complex plane $C/\{1\}$.  It has the Laurent expansion
$$\zeta(s,a)={1 \over {s-1}}+\sum_{k=0}^\infty {{(-1)^k \gamma_k(a)} \over
k!} (s-1)^k, ~~~~~~ s \neq 1. \eqno(1.2)$$
Again $\zeta(s,a)$ has a simple pole at $s=1$ with residue $1$.  By convention,
one takes $\gamma_k(1)=\gamma_k$ \cite{kreminski}.  In this paper, we are concerned with certain series representations of the constants $\gamma_k(a)$.

We let $P_1(x)=B_1(x-[x])=x-[x]-1/2$ be the first periodized Bernoulli polynomial,
and $\{x\}=x-[x]$ be the fractional part of $x$.  Addison \cite{addison} gave an
interesting series representation for $\gamma$ \cite{gerst},
$$\gamma={1 \over 2}+{1 \over 2}\sum_{n=1}^\infty n \sum_{m=2^{n-1}}^{2^n-1}{1 \over
{2m(m+1)(2m+1)}}=1-{1 \over 2}\sum_{n=1}^\infty n \sum_{m=2^{n-1}+1}^{2^n} {1 \over {m
(2m-1)}}.  \eqno(1.3)$$
His approach uses an integral representation for the zeta function in terms of $P_1$.  We note that various transformations of Addison's result are possible \cite{berndt,gerst,sandham}.  In this paper, we describe several generalizations of Addison's result.  We indicate how a family of summation representations for $\gamma_k$ may be obtained.  Representative in this regard are the first two Propositions and their proofs. 

We let $\psi=\Gamma'/\Gamma$ be the digamma function \cite{nbs}, and $H_n=\sum_{p=1}^n 1/p$ the $n$th harmonic number.  We have $H_n=\psi(n+1)+\gamma$, with
$\gamma=-\psi(1)$. 

{\bf Proposition 1}.  We have (a)
$$-\gamma_1={1 \over 4}\sum_{j=1}^\infty j \sum_{m=2^{j-1}}^{2^j-1}\left[{{(j-1)} \over 2}{{\ln 2} \over {m(m+1)(2m+1)}}-{{\ln m} \over m}+{{\ln (m+1/2)} \over {2m+1}}
-{{\ln (m+1)} \over {m+1}}\right], \eqno(1.4)$$
and (b)
$$-\gamma_1={1 \over 3}\sum_{j=1}^\infty j \sum_{m=3^{j-1}}^{3^j-1}\left\{{{(j-1)\ln 3} \over 2}\left[{1 \over {m(3m+1)}}-{1 \over {(m+1)(3m+2)}}\right] \right.$$
$$\left. -{{\ln m} \over m}+{{3\ln (m+1/3)} \over {3m+1}}+ {{3\ln (m+2/3)} \over {3m+2}}-{{\ln (m+1)} \over {m+1}}\right\}. \eqno(1.5)$$

{\bf Proposition 2}.  Let $k \geq 2$ be an integer.  Then we have
$$-\gamma_1=\sum_{j=1}^\infty j \sum_{m=k^{j-1}}^{k^j-1}\left\{ \left[{{2km+k+1} \over 
{2km(m+1)}} + \psi(km)-\psi(km+k)\right]\left(1+{{(j-1)} \over 2}\ln k\right)\right.$$
$$\left.-k \sum_{\ell=1}^k \left[{1 \over 2}\left(1-{1 \over k}\right)-{{(\ell-1)}
\over k}\right]\left[{{1+\ln\left(m+{{\ell-1} \over k}\right)} \over {km+\ell-1}}
-{{1+\ln\left(m+{{\ell} \over k}\right)} \over {km+\ell}}\right]\right\}.  \eqno(1.6)$$
Alternatively, on the right side we may write the difference
$$\psi(km)-\psi(km+k)=-\sum_{p=0}^{k-1} {1 \over {mk+p}}=H_{km-1}-H_{km+k-1}.  
\eqno(1.7)$$

Series representations that generalize Addison's approach may also be obtained
for $\gamma_k(a)$.  As an illustration, we have
{\newline \bf Proposition 3}.  Let Re $a>0$.  We have (a)
$$\gamma_0(a)=-\ln a+{1 \over {2a}}+{1 \over 4}\sum_{\nu=0}^\infty {1 \over 4^\nu}\sum_{j=0}^\infty \left.{b \over {(a+bj)(a+b+bj)(2a+b+2bj)}} \right|_{b=2^{-\nu}}$$
$$=-\psi(a),    
\eqno(1.8)$$
and (b)
$$\gamma_1(a)={{\ln a} \over {2a}}-{{\ln^2 a} \over 2}+{1 \over 4}\sum_{\nu=0}^\infty
2^{-\nu} \sum_{j=0}^\infty \left[{{\ln(a+bj)} \over {a+bj}}+{{\ln(a+b+bj)} \over {a+b+bj}}-4{{\ln[a+b(j+1/2)]} \over {2a+b+2bj}}\right]_{b=2^{-\nu}}.  \eqno(1.9)$$

Part (a) is a known result \cite{wilton2}.  However, we obtain it in a different
way.

From Proposition 3(b) we obtain
{\newline \bf Corollary 1}.  We have
$$\gamma_1\left({1 \over 2}\right)=2\gamma \ln 2+\ln^2 2 -\gamma_1.  \eqno(1.10)$$

From the proof of Proposition 3(a) we obtain
{\newline \bf Corollary 2}.  An integral representation corresponding to the 
Addison series (1.3) is given by
$$\gamma={1 \over 2}+{1 \over 2}\int_0^1 \left({{1-x} \over {1+x}}\right)\sum_{n=1}^ \infty x^{2^n-1} dx.  \eqno(1.11)$$

Our next Proposition generalizes Lemma 2.5 of \cite{berndt}.  We have
\newline{\bf Proposition 4}.  For any integer $n>1$, we have (a)
$$\gamma^2+\gamma_1=-\int_0^1\left({n \over {1-x^n}}-{1 \over {1-x}}\right)\sum_{k=1}^
\infty \left[k\ln n+\ln\left(\ln\left({1 \over x}\right)\right)\right]x^{n^k-1}dx,
\eqno(1.12)$$
(b) and for Re $a>0$, we have
$$\ln a-\psi(a)=\ln a+\gamma_0(a)=\int_0^1\left({n \over {1-x^n}}-{1 \over {1-x}}\right)\sum_{k=1}^\infty x^{an^k-1}dx, \eqno(1.13)$$ 
$$\gamma \ln a+{1 \over 2}\ln^2 a-\gamma \psi(a)+\gamma_1(a) =-\int_0^1\left({n \over {1-x^n}}-{1 \over {1-x}}\right)\sum_{k=1}^\infty \left[k\ln n+\ln\left(\ln\left({1 \over x}\right)\right)\right]x^{an^k-1}dx, \eqno(1.14)$$ 
$$\left(\gamma^2+{\pi^2 \over 6}\right)\ln a+\gamma \ln^2 a+{1 \over 3}\ln^3 a
-\left(\gamma^2+{\pi^2 \over 6}\right)\psi(a)+2\gamma \gamma_1(a)+\gamma_2(a)$$
$$=\int_0^1\left({n \over {1-x^n}}-{1 \over {1-x}}\right)\sum_{k=1}^
\infty \left[k^2\ln^2 n+2k\ln n\ln\left(\ln\left({1 \over x}\right)\right)+    \ln^2\left(\ln\left({1 \over x}\right)\right)\right]x^{an^k-1}dx.
\eqno(1.15)$$  

Our last Proposition generalizes relation (2.87) of \cite{coffey2009} for $\gamma$.
We introduce the constants (\cite{coffey2009}, Proposition 11)
$$p_{n+1}=-{1 \over {n!}}\int_0^1 (-x)_n dx={{(-1)^{n+1}} \over {n!}}\sum_{k=1}^n
{{s(k,n)} \over {k+1}}, \eqno(1.16)$$
where $(z)_n=\Gamma(z+n)/\Gamma(z)$ is the Pochhammer symbol and $s(k,\ell)$ is the
Stirling number of the first kind.  Then we have
{\newline \bf Proposition 5}.  Let Re $a>0$.  Then we have (a)
$$\ln a-\psi(a)=\ln a+\gamma_0(a)=\sum_{n=1}^\infty (n-1)! {p_{n+1} \over {(a)_n}},
\eqno(1.17)$$
and (b)
$${1 \over 2}\ln^2 a+\gamma_1(a)=\sum_{n=1}^\infty p_{n+1} \sum_{k=0}^{n-1}(-1)^k
{{n-1} \choose k} {{\ln(k+a)} \over {(k+a)}}.  \eqno(1.18)$$

From part (a) of this Proposition we obtain the following, where $_2F_1$ is the
Gauss hypergeometric function.  We have
{\newline \bf Corollary 3}.  For Re $a>0$ we have the representation
$$\ln a +\gamma_0(a)={1 \over a}\int_0^1 {{_2F_1(1,1;1+a;v)} \over {[\ln^2\left({{1-v}
\over v}\right)+\pi^2]}} {{dv} \over v}.  \eqno(1.19)$$

Recent analytic results for the Stieltjes constants may be found in \cite{coffeyjmaa} and \cite{coffey2009}, and a convenient and computationally effective asymptotic 
form in \cite{knessl}.  The work \cite{coffey2009} includes an addition 
formula for the constants together with series representations.  Although the
Stieltjes constants have been investigated for approximately a century, many open questions concerning them remain, especially including characterizing their
arithmetic properties.

We recall some relations useful in the following.

From the representation (cf. e.g., \cite{titch}, p. 14),
$$\zeta(s)={1 \over {s-1}}+{1 \over 2}-s\int_1^\infty x^{-(s+1)}P_1(x)dx, 
~~~~\mbox{Re} ~s >1, \eqno(1.20)$$
we obtain
$$\zeta^{(n)}(s)={{(-1)^n n!} \over {(s-1)^{n+1}}}+(-1)^n n\int_1^\infty {{P_1(x)}
\over x^{s+1}}\ln^{n-1} x ~dx-(-1)^n s \int_1^\infty {{P_1(x)}\over x^{s+1}}\ln^n x 
~dx.  \eqno(1.21)$$
From (1.1) we have
$$\lim_{s \to 1}\left[\zeta^{(n)}(s)-{{(-1)^n n!} \over {(s-1)^{n+1}}}\right]=(-1)^n
\gamma_n.  \eqno(1.22)$$

Equation (1.12) extends to 
$$\zeta(s,a)={a^{-s} \over 2}+{a^{1-s} \over {s-1}}-s\int_0^\infty {{P_1(x)} \over
{(x+a)^{s+1}}} dx, ~~~~\mbox{Re} ~s >-1. \eqno(1.23)$$  

We have the connection to differences of logarithmic sums
$$\gamma_j(a)-\gamma_j(b)=\sum_{n=0}^\infty \left[{{\ln^j (n+a)} \over {n+a}}-
{{\ln^j (n+b)} \over {n+b}}\right], ~~~~~~j \geq 1,  \eqno(1.24)$$
where $a, b \in C/\{-1,-2,\ldots\}$.

\medskip
\centerline{\bf Proof of Propositions}

{\it Proposition 1}.  (a)
We have from (1.12) and (1.13) the integral representation (cf. \cite{ivic}, p. 5) 
for $n \geq 1$    
$$\gamma_n=\int_1^\infty P_1(x){{\log^{n-1} x} \over x^2}(n-\log x)dx. \eqno(2.1)$$
We let $f(x)=-P_1(x)$ and put \cite{addison}
$$g_2(x)=f(x)-{1 \over 2}f(2x),  \eqno(2.2)$$
so that $g_2$ is the rectangular function
$$g_2(x) = 1/4, ~~~~0 \leq \{x\} <1/2$$
$$~~~~~~ = -1/4, ~~~~1/2 \leq \{x\} <1.  \eqno(2.3)$$
We have from (2.2)
$$\sum_{n=0}^\infty {{g_2(2^n x)} \over 2^n}=\sum_{n=0}^\infty \left[{{f(2^n x)} \over 2^n}-{{f(2^{n+1} x)} \over 2^{n+1}}\right]=f(x).  \eqno(2.4)$$
Therefore, from (2.1) we obtain the representation
$$\gamma_n=-\sum_{\nu=0}^\infty {1 \over 2^\nu}\int_1^\infty {{g_2(2^\nu x)} \over
x^2} \ln^{n-1}x (n-\ln x) dx$$
$$=-\sum_{\nu=0}^\infty \int_{2^\nu}^\infty {{g_2(y)} \over y^2}\ln^{n-1} (2^{-\nu}
y)[n-\ln(2^{-\nu}y)]dy, \eqno(2.5)$$
with the change of variable $y=2^\nu x$.  We write the integral over $y$ as
$\int_{2^\nu}^\infty dy=\sum_{j=\nu}^\infty \int_{2^j}^{2^{j+1}}dy$ and then
interchange the resulting double sum in (2.5).  Therefore, we obtain
$$\gamma_n=-\sum_{j=0}^\infty \sum_{\nu=0}^j \int_{2^j}^{2^{j+1}} {{g_2(y)} \over y^2}\ln^{n-1} (2^{-\nu} y)[n-\ln(2^{-\nu}y)]dy. \eqno(2.6)$$

We now specialize to $n=1$:
$$\gamma_1=-\sum_{j=0}^\infty \sum_{\nu=0}^j \int_{2^j}^{2^{j+1}} {{g_2(y)} \over y^2}[1-\ln(2^{-\nu}y)]dy$$
$$=-\sum_{j=0}^\infty \sum_{\nu=0}^j \int_{2^j}^{2^{j+1}} {{g_2(y)} \over y^2}[1+\nu \ln 2 -\ln y]dy$$
$$=-\sum_{j=0}^\infty (j+1)\int_{2^j}^{2^{j+1}} {{g_2(y)} \over y^2}[1+{j \over 2}
\ln 2 -\ln y]dy.  \eqno(2.7)$$
We then use the integral
$$\int_m^{m+1} {{g_2(y)} \over y^2}\ln y ~dy={1 \over 4}\left(\int_m^{m+1/2}-
\int_{m+1/2}^{m+1}\right){{\ln y} \over y^2}dy$$
$$={1 \over 4}\left[{{1+\ln m} \over m}
-{{4(1+\ln(m+1/2))} \over {2m+1}}+{{1+\ln(m+1)} \over {m+1}}\right].  \eqno(2.8)$$
We find
$$-\gamma_1={1 \over 4}\sum_{j=1}^\infty j \sum_{m=2^{j-1}}^{2^j-1} \left[{{1+{{j-1}
\over 2}\ln 2} \over {m(m+1)(2m+1)}} -{{1+\ln m} \over m}
+{{4(1+\ln(m+1/2))} \over {2m+1}}-{{1+\ln(m+1)} \over {m+1}}\right].  \eqno(2.9)$$
With the use of partial fractions on the first term on the right side of this
equation, we obtain the form (1.4).

For part (b), we proceed similarly, employing the function
$$g_3(x)=f(x)-{1 \over 3}f(3x)= 1/3, ~~~~0 \leq \{x\} <1/3,$$
$$~~~~~~~~~~~~~~~~~~~~~~~~~~ = 0, ~~~~1/3 \leq \{x\} <2/3, $$
$$~~~~~~~~~~~~~~~~~~~~~~~~~ = -1/3, ~~~~2/3 \leq \{x\} <1,  \eqno(2.10)$$
so that $\sum_{n=0}^\infty {{g_3(3^n x)} \over 3^n}=f(x)$.  We use the integrals
$$\int_m^{m+1} {{g_3(y)} \over y^2}dy={1 \over 3}\left(\int_m^{m+1/3}-
\int_{m+2/3}^{m+1}\right){{dy} \over y^2}$$
$$={1 \over 3}\left[{1 \over {m(3m+1)}}-{1 \over {(m+1)(3m+2)}}\right], \eqno(2.11a)$$
and
$$\int_m^{m+1} {{g_3(y)} \over y^2}\ln y ~dy={1 \over 3}\left(\int_m^{m+1/3}-
\int_{m+2/3}^{m+1}\right){{\ln y} \over y^2}dy$$
$$={1 \over 3}\left[{{1+\ln m} \over m}-{{3(1+\ln(m+1/3)} \over {3m+1}}
-{{3(1+\ln(m+2/3)} \over {3m+2}}+{{1+\ln(m+1)} \over {m+1}}\right].  \eqno(2.11b)$$
Again, an application of partial fractions gives the final form (1.5).

{\it Proposition 2}.
Proposition 1 is further extended with the use of functions for integers 
$k \geq 2$
$$g_k(x)=f(x)-{1 \over k}f(kx),  \eqno(2.12)$$
with $\sum_{n=0}^\infty {{g_k(k^n x)} \over k^n}=f(x)$.  Since $f$ is periodic,
$g_k$ is also periodic of period $1$.  From the expression
$$g_k(x)={1 \over 2}\left(1-{1 \over k}\right)+{1 \over k}\{kx\}-\{x\}, \eqno(2.13)$$
we obtain the values of $g_k(x)$ on subintervals $\left[{{j-1} \over k},{j \over k}
\right)$ for $j=1,2,\ldots,k$.  We have
$$g_k(x)={1 \over 2}\left(1-{1 \over k}\right)-{{(j-1)} \over k}, ~~~~~~x \in \left[{{j-1} \over k},{j \over k}\right).  \eqno(2.14)$$
This enables the development of summation expressions for any $\gamma_n$ as a result 
of integrations over a function $g_k(x)$.  We have from (2.1)
$$\gamma_n=-\sum_{\nu=0}^\infty {1 \over k^\nu}\int_1^\infty {{g_k(k^\nu x)} \over
x^2} \ln^{n-1}x (n-\ln x) dx$$
$$=-\sum_{\nu=0}^\infty \int_{k^\nu}^\infty {{g_k(y)} \over y^2}\ln^{n-1} (k^{-\nu}
y)[n-\ln(k^{-\nu}y)]dy$$
$$=-\sum_{j=0}^\infty \sum_{\nu=0}^j \int_{k^j}^{k^{j+1}} {{g_k(y)} \over y^2}\ln^{n-1} (k^{-\nu} y)[n-\ln(k^{-\nu}y)]dy. \eqno(2.15)$$

So at $n=1$ we obtain
$$-\gamma_1=\sum_{j=0}^\infty \sum_{\nu=0}^j \int_{k^j}^{k^{j+1}} {{g_k(y)} \over y^2}[1-\ln(k^{-\nu}y)]dy$$
$$=\sum_{j=0}^\infty (j+1)\int_{k^j}^{k^{j+1}} {{g_k (y)} \over y^2}[1+{j \over 2}
\ln k -\ln y]dy.  \eqno(2.16)$$
Now using the piecewise values in (2.14), we have
$$\int_m^{m+1} {{g_k(y)} \over y^2}dy=\sum_{\ell=1}^k \left[{1 \over 2}\left(1-{1 
\over k}\right)-{{(\ell-1)} \over k}\right] \int_{m+(\ell-1)/k}^{m+\ell/k} {{dy}
\over y^2}$$
$$=\sum_{\ell=1}^k \left[{1 \over 2}\left(1-{1 \over k}\right)-{{(\ell-1)} \over k}\right] {k \over {(km+\ell)(km+\ell-1)}}$$
$$={{2km+k+1} \over {2km(m+1)}} +\psi(km)-\psi(km+k).  \eqno(2.17)$$
We also have
$$\int_m^{m+1} {{g_k(y)} \over y^2}\ln y ~dy=\sum_{\ell=1}^k \left[{1 \over 2}\left(1-{1 \over k}\right)-{{(\ell-1)} \over k}\right] \int_{m+(\ell-1)/k}^{m+\ell/k} \ln y {{dy} \over y^2}$$
$$=\sum_{\ell=1}^k \left[{1 \over 2}\left(1-{1 \over k}\right)-{{(\ell-1)} \over k}\right] \left[{{1+\ln\left(m+{{\ell-1} \over k}\right)} \over {km+\ell-1}}
-{{1+\ln\left(m+{{\ell} \over k}\right)} \over {km+\ell}}\right].  \eqno(2.18)$$
combining (2.16), (2.17), and (2.18), we arrive at the Proposition.

{\it Proposition 3}.  (a)  From (1.13) and (1.2) we have
$$\lim_{s \to 1}\left[\zeta(s,a)-{a^{1-s} \over {s-1}}\right]=-\ln a+{1 \over {2a}}
-\int_0^\infty {{P_1(x)} \over {(x+a)^2}} dx, \eqno(2.19)$$
so that
$$\gamma_0(a)=-\ln a+{1 \over {2a}}+\int_0^\infty {{f(x)} \over {(x+a)^2}} dx. \eqno(2.20)$$
Using the functions $g_k$ of (2.12) we have
$$\int_0^\infty {{f(x)} \over {(x+a)^2}} dx=\sum_{\nu=0}^\infty {1 \over k^\nu}
\int_0^\infty {{g_k(k^\nu x)} \over {(x+a)^2}}dx$$
$$=\sum_{\nu=0}^\infty {1 \over k^{2\nu}}\sum_{j=0}^\infty \int_j^{j+1} {{g_k(y)}
\over {(k^{-\nu}y+a)^2}}dy.  \eqno(2.21)$$
We now specialize to $k=2$ and obtain
$$\int_0^\infty {{f(x)} \over {(x+a)^2}} dx=\sum_{\nu=0}^\infty {1 \over 4^\nu}
\sum_{j=0}^\infty \int_j^{j+1} {{g_2(y)}\over {(2^{-\nu}y+a)^2}}dy$$
$$={1 \over 4}\sum_{\nu=0}^\infty {1 \over 4^\nu}\sum_{j=0}^\infty \left(\int_j^{j+1/2}
-\int_{j+1/2}^{j+1}\right){{dy} \over {(2^{-\nu}y+a)^2}}$$
$$={1 \over 4}\sum_{\nu=0}^\infty {1 \over 4^\nu}\sum_{j=0}^\infty \left. {b \over {(a+bj)(a+b+bj)(2a+b+2bj)}}\right|_{b=2^{-\nu}}$$
$$={1 \over 4}\sum_{\nu=0}^\infty \left[2\psi\left({{2a+b} \over {2b}}\right)-
\psi\left({a \over b}\right)-\psi\left({{a+b} \over b}\right)\right]_{b=2^{-\nu}}
.  \eqno(2.22)$$
We now apply the functional equation and duplication formula for the digamma
function:
$$\int_0^\infty {{f(x)} \over {(x+a)^2}} dx={1 \over 4}\sum_{\nu=0}^\infty \left[2\psi\left({{2a+b} \over {2b}}\right)-2\psi\left({a \over b}\right)-{b \over a}\right]_{b=2^{-\nu}}$$
$$= {1 \over 2}\sum_{\nu=0}^\infty \left[\psi\left({a \over b}+{1 \over 2}\right)-\psi\left({a \over b}\right) \right]_{b=2^{-\nu}} -{1 \over {2a}}$$
$$=\sum_{\nu=0}^\infty \left[\psi\left({{2a} \over b}\right)-\psi\left({a \over b}\right) -\ln 2\right]_{b=2^{-\nu}} -{1 \over {2a}}.  \eqno(2.23)$$
We employ a standard integral representation for $\psi$ (\cite{grad}, p. 943) to 
write
$$\sum_{\nu=0}^\infty \left[\psi\left({{2a} \over b}\right)-\psi\left({a \over b}\right) -\ln 2\right]_{b=2^{-\nu}}=\lim_{T \to \infty}\sum_{\nu=0}^T \left[
\int_0^1 {{(t^{2a2^\nu -1} -t^{a2^\nu -1})} \over {t-1}}dt-\ln 2\right]$$
$$=\lim_{T \to \infty}\left[\int_0^1 {{(t^{a2^{T+1}-1}-t^{a-1})} \over {t-1}}dt
-(T+1)\ln 2\right]$$
$$=\lim_{T \to \infty}\left[\psi(a2^{T+1})-\psi(a)-(T+1)\ln 2\right]$$
$$=-\psi(a)+\ln a, \eqno(2.24)$$
where the above interchange of summation and integration is justified by absolute
convergence of the integral, and where in the last step we applied the large argument asymptotic form for $\psi$ (\cite{nbs}, p. 259).
Therefore, from (2.20), (2.23), and (2.24), we find (1.6).

(b) Using (1.13) and (1.2) we have
$$\lim_{s \to 1}\left[\zeta'(s,a)-a^{1-s}\left({1 \over {(s-1)^2}}+{{\ln a} \over  {s-1}}\right)\right]={{\ln^2 a} \over 2}-{{\ln a} \over {2a}}
-\int_0^\infty {{P_1(x)} \over {(x+a)^2}}[1-\ln(x+a)]dx, \eqno(2.25)$$
giving
$$\gamma_1(a)=-{{\ln^2 a} \over 2}+{{\ln a} \over {2a}}+\int_0^\infty {{P_1(x)} \over {(x+a)^2}}[1-\ln(x+a)]dx. \eqno(2.26)$$
With $k=2$ in (2.12) we have
$$-\int_0^\infty {{P_1(x)} \over {(x+a)^2}}[1-\ln(x+a)]dx
=\sum_{\nu=0}^\infty {1 \over 4^\nu} \sum_{j=0}^\infty \int_j^{j+1} {{g_2(y)} \over {(2^{-\nu}y+a)^2}}[1-\ln(2^{-\nu}y+a)] dy$$
$$={1 \over 4}\sum_{\nu=0}^\infty {1 \over 4^\nu}\sum_{j=0}^\infty \left(\int_j^{j+1/2}
-\int_{j+1/2}^{j+1}\right)[1-\ln(2^{-\nu}y+a)]{{dy} \over {(2^{-\nu}y+a)^2}}.  
\eqno(2.27)$$
We then carry out the integrations and simplify.  Using (2.25) completes part (b).

{\it Corollary 1}.  We put $a=1/2$ in (1.9) and manipulate the resulting series to
write
$$\gamma_1(1/2)+\ln 2+{1 \over 2}\ln^2={{\ln(1/2)} \over 4}\sum_{\nu=0}^\infty
2^{-\nu+1}\sum_{j=0}^\infty \left[{1 \over {1+2^{-\nu+1}j}}+{1 \over {1+2^{-\nu+1}(j+1)}}-{2 \over {1+2^{-\nu+1}(j+1/2)}}\right]$$
$$+{1 \over 4}\sum_{\nu=0}^\infty 2^{-\nu}\sum_{j=0}^\infty \left[{{\ln(1+bj)} \over
{1+bj}}+{{\ln[1+b(j+1)]} \over {1+b(j+1)}}-2{{\ln[1+b(j+1/2)]} \over
{1+b(j+1/2)}}\right]_{b=2^{-\nu}}$$
$$+{1 \over 2}\sum_{j=0}^\infty \left[{{\ln(1+2j)} \over
{1+2j}}+{{\ln[1+2(j+1)]} \over {1+2(j+1)}}-2{{\ln[1+2(j+1/2)]} \over
{1+2(j+1/2)}}\right].  \eqno(2.28)$$
By the method of (2.23)-(2.24), the first line on the right side of this equation
has the value $-\ln 2(\gamma+\ln 2-1)$.  From Proposition 3(b) at $a=1$, the
second line on the right side of (2.28) is $\gamma_1$.  The following evaluates
the third line of the right side of (2.28).

{\bf Lemma 1}.  We have
$${1 \over 2}\sum_{j=0}^\infty \left[{{\ln(1+2j)} \over
{1+2j}}+{{\ln[1+2(j+1)]} \over {1+2(j+1)}}-2{{\ln[1+2(j+1/2)]} \over
{1+2(j+1/2)}}\right]={1 \over 2}\left[\gamma_1\left({1 \over 2}\right)-\gamma_1\right]
+\ln^2 2.  \eqno(2.29)$$
{\it Proof}.  We write
$${1 \over 4}\sum_{j=0}^\infty \left[{{\ln[2(j+1/2)]} \over
{j+1/2}}+{{\ln[2(j+3/2)]} \over {j+3/2}}-2{{\ln[2(j+1)]} \over
{j+1}}\right]={{\ln 2} \over 4}\sum_{j=0}^\infty \left[{1 \over {j+1/2}}+{1 \over {j+3/2}}-{2 \over {j+1}}\right]$$
$$+{1 \over 4}\left[\gamma_1\left({1 \over 2}\right)-\gamma_1
+\gamma_1\left({3 \over 2}\right)-\gamma_1\right], \eqno(2.30)$$
where we used (1.13) at $j=1$.  Then the sum of (2.30) is given by
$${{\ln 2} \over 2}(2\ln 2-1)+{1 \over 4}\left[\gamma_1\left({1 \over 2}\right)+\gamma_1\left({3 \over 2}\right)-2\gamma_1\right]$$
$$={{\ln 2} \over 2}(2\ln 2-1)+{1 \over 2}\left[\gamma_1\left({1 \over 2}\right)-\gamma_1+\ln 2\right]$$
$$={1 \over 2}\left[\gamma_1\left({1 \over 2}\right)-\gamma_1\right]
+\ln^2 2.  \eqno(2.31)$$
In obtaining this result we have used $\gamma_k(a+1)=\gamma_k(a)-(\ln^k a)/a$,
that follows easily from the series form of $\zeta(s,a)$, at $k=1$.

Then combining the terms on the left and right sides of (2.28) yields (1.10).

{\it Corollary 2}.  We have, by applying for instance Theorem 2.2 of \cite{berndt},
$$\int_0^1 \left({{1-x} \over {1+x}}\right)x^p dx=-{1 \over {p+1}}+H_{p/2}-H_{(p-1)/2}
, ~~~~~~\mbox{Re} ~p >-1.  \eqno(2.32)$$
Then, with the interchange of the summation and integration justified by the absolute
convergence of the integral, we have
$$\int_0^1 \left({{1-x} \over {1+x}}\right) \sum_{n=1}^\infty x^{2^n-1} dx=
-1+\sum_{n=1}^\infty [H_{p/2}-H_{(p-1)/2}]_{p=2^n-1}, \eqno(2.33)$$
giving
$${1 \over 2}+{1 \over 2}\int_0^1 \left({{1-x} \over {1+x}}\right) \sum_{n=1}^\infty x^{2^n-1} dx={1 \over 2}\sum_{n=1}^\infty \left[\psi\left({{p+1}
\over 2}\right)-\psi\left({p \over 2}+1\right)\right]_{p=2^n-1}.  \eqno(2.34)$$
By the method of (2.22)-(2.24), the Corollary follows.

{\it Proposition 4}.  Part (a) is based upon the integral
$$\int_0^1\left({1 \over {\ln x}}+{1 \over {1-x}}\right)\ln^s\left({1 \over x}\right)
dx=\Gamma(s)[s\zeta(s+1)-1],  ~~~~~~\mbox{Re} ~s>-1.  \eqno(2.35)$$
This equation follows easily by term-by-term integration, as we have
$$\int_0^1\left({1 \over {\ln x}}+{1 \over {1-x}}\right)\ln^s\left({1 \over x}\right)
dx=\int_0^1 {{\ln^s(1/x)} \over {\ln x}}dx+\sum_{j=0}^\infty \int_0^1 x^j \ln^s\left(
{1 \over x}\right)dx$$
$$=-\Gamma(s)+\Gamma(s+1)\sum_{j=0}^\infty {1 \over {(j+1)^{s+1}}}, \eqno(2.36)$$
where we used the change of variable $u=-\ln x$.
We apply Lemma 2.4 of \cite{berndt} that we write in the following form.  For any
integer $n>1$ and Re $x>0$, we have
$${1 \over {\ln x}}+{1 \over {1-x}}=\sum_{k=1}^\infty {{\sum_{j=1}^{n-1} (n-j)x^{(j-1)/n^k}} \over {n^k \sum_{j=0}^{n-1} x^{j/n^k}}}.  \eqno(2.37)$$
We then proceed as in the proof of Lemma 2.5 of \cite{berndt} to find
$$\int_0^1\left({1 \over {\ln u}}+{1 \over {1-u}}\right)\ln^s\left({1 \over u}\right)
du=\int_0^1 \left({n \over {1-x^n}}-{1 \over {1-x}}\right)\ln^s\left({1 \over x}\right
)\sum_{k=1}^\infty n^{ks} x^{n^k-1}dx.  \eqno(2.38)$$
We perform $\left.{\partial \over {\partial s}}\right|_{s=0}$ on (2.35) and (2.38),
and (1.12) follows.

Part (b) is based upon the integral
$$\int_0^1\left({1 \over {\ln x}}+{1 \over {1-x}}\right)x^{a-1}\ln^s\left({1 \over x}\right)dx=\Gamma(s)[s\zeta(s+1,a)-a^{-s}],  ~~~~~~\mbox{Re} ~s>-1, ~~\mbox{Re} ~a
>0.  \eqno(2.39)$$
Using Lemma 2.4 of \cite{berndt} we find
$$\int_0^1\left({1 \over {\ln u}}+{1 \over {1-u}}\right)u^{a-1}\ln^s\left({1 \over u}\right) du=\int_0^1 \left({n \over {1-x^n}}-{1 \over {1-x}}\right)\ln^s\left({1 \over x}\right)\sum_{k=1}^\infty n^{ks} x^{an^k-1}dx.  \eqno(2.40)$$
Taking $s=0$ in (2.38) and (2.39) gives (1.13).  Performing $\left.\left({\partial \over {\partial s}}\right)^j\right|_{s=0}$, $j=1,2$ on (2.39) and (2.40),
respectively gives (1.14) and (1.15).

{\it Proposition 5}.  We employ the generating function (\cite{coffey2009}, (2.78))
$$\sum_{n=1}^\infty p_{n+1}z^{n-1} ={1 \over z}+{1 \over {\ln(1-z)}}, ~~~~~~|z|<1,
\eqno(2.41)$$
so that by (2.39) we have
$$\Gamma(s)[s\zeta(s+1,a)-a^{-s}]=\sum_{n=1}^\infty p_{n+1}\int_0^1 x^{a-1}(1-x)^{n-1}
\ln^s \left({1 \over x}\right)dx.  \eqno(2.42)$$
For part (a), we take $s=0$ in this equation and evaluate the integral using the
Beta function.  For part (b), we differentiate both sides of (2.42) with respect
to $s$ and put $s=0$.  We find
$$-\gamma \ln a-{1 \over 2}\ln^2 a+\gamma \psi(a)-\gamma_1(a)=\sum_{n=1}^\infty
p_{n+1} \int_0^1 x^{a-1} (1-x)^{n-1} \ln\left(\ln\left({1 \over x}\right)\right)dx.
\eqno(2.43)$$
By logarithmic differentiation we obtain the integral for Re $s>-1$
$$\int_0^1 x^k \ln^s\left({1 \over x}\right)\ln\left(\ln\left({1 \over x}\right) \right)dx={{\Gamma(s+1)} \over {(k+1)^{s+1}}}[\psi(s+1)-\ln(k+1)],  \eqno(2.44)$$
giving at $s=0$
$$\int_0^1 x^k \ln\left(\ln\left({1 \over x}\right) \right)dx=-{{[\gamma+\ln(k+1)]}
\over {k+1}}.  \eqno(2.45)$$
Then by binomial expansion we have
$$\int_0^1 x^{a-1} (1-x)^{n-1} \ln\left(\ln\left({1 \over x}\right)\right)dx
=-\sum_{k=0}^{n-1} (-1)^k {{n-1} \choose k} {{[\gamma+\ln(k+a)]}
\over {k+a}}$$
$$=-\gamma {{(n-1)!} \over {(a)_n}} -\sum_{k=0}^{n-1} (-1)^k {{n-1} \choose k} {{\ln(k+a)} \over {k+a}}.  \eqno(2.46)$$
By using (2.43) and (1.17) we find (1.18).

For Corollary 3 we apply the integral representation (2.85) of \cite{coffey2009}
for $p_{n+1}$, so that by (1.17)
$$\ln a+\gamma_0(a)=\sum_{n=1}^\infty {{(n-1)!} \over {(a)_n}}\int_0^\infty {{du}
\over {(1+u)^n (\ln^2 u+\pi^2)}}$$
$$={1 \over a}\int_0^\infty {1 \over {(\ln^2 u+\pi^2)}}{1 \over {(1+u)}} ~_2F_1
\left(1,1;1+a;{1 \over {1+u}}\right)du.  \eqno(2.47)$$

{\bf Remarks}.  From (2.6) we have for $n \geq 1$
$$-\gamma_n=\sum_{j=0}^\infty \sum_{\nu=0}^j \int_{2^j}^{2^{j+1}} {{g_2(y)} \over y^2}\sum_{\ell=0}^{n-1} {{n-1} \choose \ell} (-\nu \ln 2)^\ell \ln^{n-\ell-1} y[n+\nu\ln 2-\ln y]dy, \eqno(2.48)$$
and more generally from (2.15) for $k \geq 2$ we have
$$-\gamma_n=\sum_{j=0}^\infty \sum_{\nu=0}^j \int_{k^j}^{k^{j+1}} {{g_k(y)} \over y^2}\sum_{\ell=0}^{n-1} {{n-1} \choose \ell} (-\nu \ln k)^\ell \ln^{n-\ell-1} y[n+\nu\ln k-\ln y]dy. \eqno(2.49)$$

With (2.20) and (1.8) we have recovered a known integral representation for 
$\ln \Gamma$ or the polygamma functions $\psi^{(j)}$ in terms of $P_1$ (e.g.,
\cite{edwards}, Section 6.3).  However, our route has been very different from
say the use of Euler-Maclaurin summation.

The result (1.10) also follows from the relation $\zeta(s,1/2)=(2^s-1)\zeta(s)$.  
The method of Corollary 1 shows an instance of the functional relation 
contained in (1.9).

In Proposition 5(a), when $a=1$, we have $(1)_n=n!$, and we recover (2.87) of
\cite{coffey2009}.  
Other forms of part (b) may be written by using the integral representation
$$\sum_{k=0}^{n-1} (-1)^k {{n-1} \choose k}{{\ln(k+a)} \over {k+a}}=-\int_0^1 \left[
{{(n-1)!} \over {(a)_n}} -{u^{a-1} \over a} ~_2F_1(1,1-n;1+a;u)\right]{{du} \over
{\ln u}}.  \eqno(2.50)$$
This expression has been obtained by inserting an integral representation for 
$\ln$ on the left side, interchanging summation and integration, and making a 
change of variable.

We note that representations of other constants may be obtained from Lemma 2.4 of 
\cite{berndt}.  We let li$(z)\equiv \int_0^z dt/\ln t$ be the logarithmic integral.
Then we have
{\newline \bf Corollary 4}.  For any integer $n>1$ we have
$$-\ln(1-y)+\mbox{li}(y)=\int_0^y \left({n \over {1-x^n}}-{1 \over {1-x}}\right)
\sum_{k=1}^\infty x^{n^k-1} dx.  \eqno(2.51)$$

Our results for $\gamma_k$ also imply representations for the coefficients
$\eta_j$ that appear in the Laurent expansion of $\zeta'/\zeta$ about $s=1$
(e.g., \cite{coffeyprsa}).  As an example, we have $\eta_1=\gamma^2+2\gamma_1$.

\bigskip
\centerline{\bf Summary}
\medskip

We have shown that the method of Addison may be generalized in several
directions.  It applies not only to the Euler constant $\gamma$, but to
the Stieltjes constants $\gamma_k$ for the Riemann zeta function.  It 
further extends to $\gamma_k(a)$ for the Hurwitz zeta function.  Moreover,
one may develop a set of parameterized series representations by using the
step-wise functions $g_k(x)$.  On yet a broader scale, the series developments
are applicable any time the first periodic Bernoulli polynomial $P_1$ is
integrated, including in numerous occurrences of Euler-Maclaurin summation.

With Proposition 4, we have generalized representations of $\gamma$ given by
Ramanujan and Berndt and Bowman \cite{berndt}.



\pagebreak

\end{document}